\documentclass[sigconf]{acmart}

\makeatletter
\def\@ACM@checkaffil{
    \if@ACM@instpresent\else
    \ClassWarningNoLine{\@classname}{No institution present for an affiliation}%
    \fi
    \if@ACM@citypresent\else
    \ClassWarningNoLine{\@classname}{No city present for an affiliation}%
    \fi
    \if@ACM@countrypresent\else
        \ClassWarningNoLine{\@classname}{No country present for an affiliation}%
    \fi
}
\makeatother

\AtBeginDocument{%
  \providecommand\BibTeX{{%
    \normalfont B\kern-0.5em{\scshape i\kern-0.25em b}\kern-0.8em\TeX}}}

\acmConference[CHI In2Writing Workshop  '23]{Make sure to enter the correct
  conference title from your rights confirmation emai}{Apr 23,
  2023}{Hamburg, Germany}
\pdfoutput=1
\usepackage{multirow}
\usepackage{colortbl}
\usepackage{xspace}
\usepackage{pifont}
\usepackage{makecell}

\setcopyright{none}
\renewcommand\footnotetextcopyrightpermission[1]{}
\begin{document}




\title{{\fontsize{17pt}{17.6pt}\selectfont \system: Evaluating Interactive Human-LM Co-writing Systems}}

%
%
%


\author{Hua Shen}
\email{huashen218@psu.edu}
\affiliation{%
  \institution{Pennsylvania State University, USA}
  }

\author{Tongshuang Wu}
\email{sherryw@cs.cmu.edu}
\affiliation{%
  \institution{Carnegie Mellon University, USA}
  }


\newcommand{\system}{\textsl{\texttt{Parachute}}\xspace}

\newcommand{\kenneth}[1]{{\small\textcolor{blue}{\bf [#1 --Ken]}}}
\newcommand{\sherry}[1]{{\small\textcolor{violet}{\bf [#1 --Sherry]}}}
\newcommand{\hua}[1]{{\small\textcolor{orange}{\bf [#1 --Hua]}}}

\newcommand{\eg}{\emph{e.g.,}\xspace}%
\newcommand{\ie}{\emph{i.e.,}\xspace}

\newcommand*{\img}[1]{%
    \raisebox{-.1\baselineskip}{%
        \includegraphics[
        height=.8\baselineskip,
        width=.8\baselineskip,
        keepaspectratio,
        ]{#1}%
    }%
}

\begin{abstract}

A surge of advances in language models (LMs) has led to significant interest in using LMs to build co-writing systems, in which humans and LMs interactively contribute to a shared writing artifact.
However, there is a lack of studies assessing co-writing systems in interactive settings.
We propose a human-centered evaluation framework, \system, for interactive co-writing systems. 
\system showcases an integrative view of interaction evaluation, where each evaluation aspect consists of categorized practical metrics.
%
%
Furthermore, we present \system with a use case to demonstrate how to evaluate and compare co-writing systems using \system.
%
\vspace{-0pt}

%
%
%
%
%
%
%
%
%

\end{abstract}

\begin{teaserfigure}
    \centering
  \includegraphics[width=.9\textwidth]{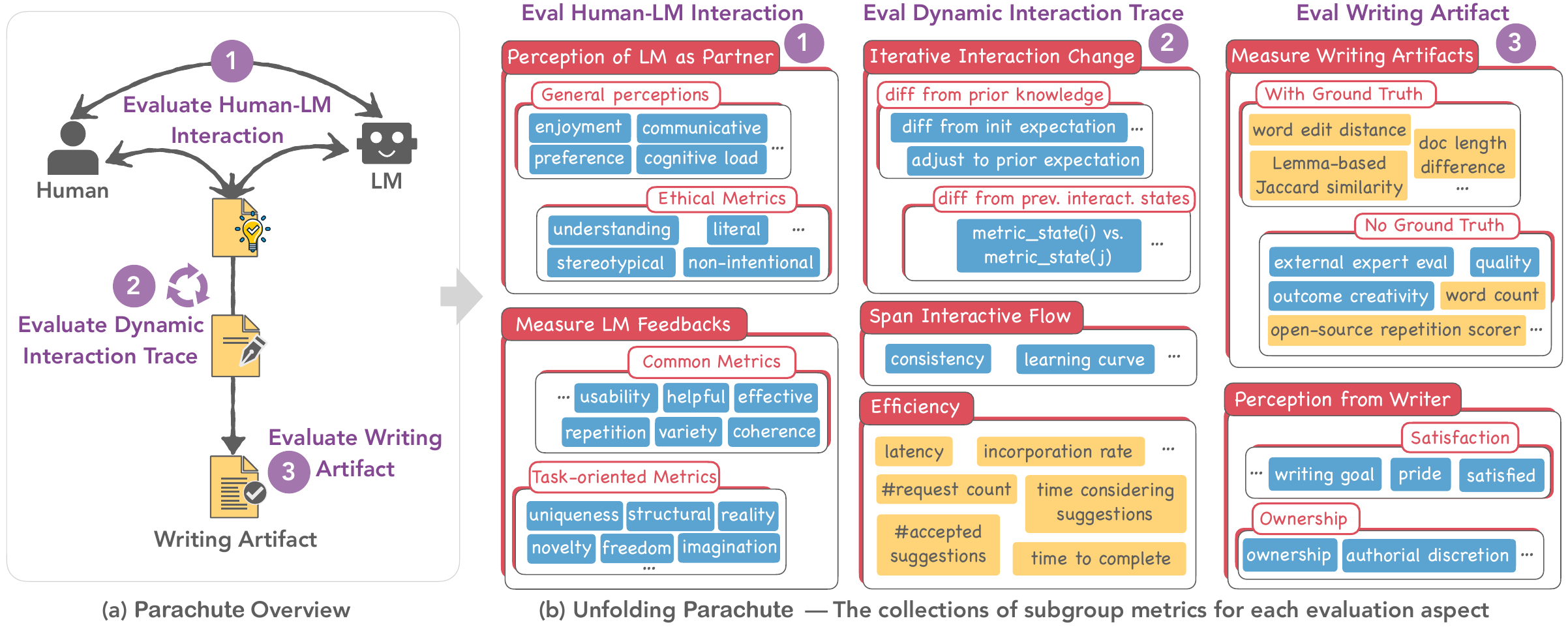}
  \vspace{-.5em}
  \caption{
   \system: human-centered integrative evaluations on interactive co-writing systems. 
   (a) \system overview. (b) The subgroup metrics for each evaluation aspect.
   ( \colorbox[HTML]{FECC66}{\color[HTML]{4B4D4C}{\texttt{yellow}}} and \colorbox[HTML]{4C91C4}{\color[HTML]{FFFFFF}{\texttt{blue}}} indicates objective and subjective metrics, respectively.)
  }
  \label{fig:framework}
\end{teaserfigure}



\maketitle

\vspace{-3pt}
\section{ Introduction}
\label{sec:introduction}

Language models (LMs) have advanced significantly, showcasing previously unheard-of capabilities in solving a wide spectrum of generation and language understanding tasks~\cite{rae2021scaling,palm,ouyang2022training}.
This has spurred great academic and public interest in using LMs to build writing assistants, in which humans collaborate with LMs to 
paraphrase sentences (\eg \href{https://quillbot.com/}{QuillBot}), autocomplete sentences~\cite{chen2019gmail},  write stories~\cite{akoury-etal-2020-storium}, etc.
Despite the interactive co-writing process, the co-writing systems are at present primarily tested in non-interactive settings~\cite{wordcraft,beyondtextgeneration}.
%
%
Specifically,
current studies commonly conduct evaluations only on the final, co-written article~\cite{dramatron}, or prior- and post-human assessment on perceiving LMs~\cite{wordcraft,integrativeheaps}, etc.
Consequently, these evaluations fail to capture the \emph{delta} (or dynamic shift) of human-LM interactions. 
Consider, for instance, a scientific paper writing task involving two types participants, freshmen and a university professors. 
While it may not be surprising that the professor-LM team achieves significantly better writing quality, this metric does not reflect the fact the impact of the LM on its users: 
Freshmen might have benefited significantly from the model in terms of the scientific paper structure, grammar correction, etc., whereas professors might have achieved similar writing performance even without the LM.
%
%
%
%
In other words, the quality of the final article cannot genuinely reflect the co-writing system's influence on users.
Instead, we should assess users' dynamic interaction improvements to indicate system capability, such as
the relative quality change between two iterated articles from the same user.


%
%
%
%

In spite of the importance, to the best of our knowledge, there are few studies that examine \emph{how to assess the interaction shift in the iterative co-writing process}, and \emph{how to depict an integrative view for evaluating interactive co-writing systems.}
%
%
%
%
%
To close this gap, we propose \system: a human-centered evaluation framework for human-LM interactive co-writing systems.
We identify the key components and interaction aspects for evaluating the co-writing systems.
We further collect the comprehensive types of evaluation metrics under each aspect, 
including the novel measurements designed for dynamic interaction assessments, and the other two conventional yet important aspects (\ie human-LM interaction and writing artifact evaluation).
%
%
Though a case study, we show that \system can be effectively used as a thinking tool for comprehensively evaluating and comparing the co-writing systems.

\vspace{-3pt}
\section{\system Framework}
\label{sec:approach}

%
%
%
%
We aim to propose \system as a guideline that can assist researchers in evaluating and analyzing interactive co-writing systems more comprehensively and fairly.

To this end, we focus on identifying key axes for co-writing systems.
To begin with, grounded on the Co-Creative Framework for Interaction Design (COFI) model~\cite{rezwana2022designing}, we identify three key \textbf{components} in human-LM co-writing systems: 
it involves both  {\textbf{human}} and  {\textbf{LM}} collaborating on a shared  {\textbf{writing artifact}} (\eg essay, story, paper, etc.) as partners~\cite{rezwana2022designing}.
%
%
%
Among these three components, \emph{humans} are the primary decision-makers for interacting with LMs and writing artifacts.
To reflect its importance, we design \system to be a human-centered evaluation framework, meaning that the ultimate objectives of these metrics is help humans achieve their various writing needs (\eg, better user experience, higher-quality writing artifact, etc.).
%

As shown in Figure~\ref{fig:framework}, \system further recognizes three \textbf{aspects}, which are important around these components and also the interactions between them,
including 
\hspace*{.0em}\img{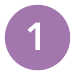}\hspace*{-.1em}
{\emph{evaluate human-LM interaction}}, 
\hspace*{.0em}\img{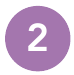}\hspace*{-.1em}
{\emph{evaluate dynamic interaction trace}}, and 
\hspace*{.0em}\img{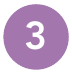}\hspace*{-.1em}
{\emph{evaluate writing artifact}}.
%
%
%
Based on existing evaluation approaches, which might capture insufficient evaluation aspects to reflect the system capability holistically~\cite{wordcraft, dramatron, beyondtextgeneration} or focus much on one-time metrics that neglect dynamic interaction changes~\cite{integrativeheaps,lee2022evaluating},
\system seeks to contribute in two folds. First, \system presents a more integrative view of interaction evaluation supported by practical metrics. Besides, \system explicitly extracts a set of metrics to assess the dynamic change along iterations.

\vspace{-3pt}
\section{{\system to Practical Metrics}}
\label{sec:metrics}

How can \system guide practical evaluation? 
We use \system to analyze existing evaluations,  and reflect on what axes these metrics emphasize
%
Concretely, we use an inductive approach to collect the practical metrics adopted in
%
%
%
state-of-the-art co-writing systems (\eg \emph{Wordcraft}~\cite{wordcraft}, \emph{Integrative Leaps}~\cite{integrativeheaps}, \emph{Beyond Text Generation}~\cite{beyondtextgeneration}, \emph{Dramatron}~\cite{dramatron}, etc.)
Then we categorize them into subgroups and fit into \system framework.
Figure~\ref{fig:framework}(b) depicts the categorized evaluation metrics for each \system aspect.
We next clearly define the aspects and metric subgroups in \system, and briefly explain the underlying motivation.
Please see Table~\ref{tab:framework} in \textbf{Appendix~\ref{appdx:metrics}} for more elaborated metrics and details.\footnote{Note that we do not aim to build enumerated lists of evaluation metrics. Instead, we focus on introducing the motivation and process of creating these evaluation aspects and subgroups, which can be generalized in broader use.}.

\vspace{2pt}
\noindent \hspace*{.0em}\img{figures/num1.pdf}\hspace*{-.1em} \textbf{Evaluating Human-LM interaction.} These metrics measure interactions between the co-writers (\ie human \& LM).
%
%
%
%
%
%
%
Suppose humans perceive LM as a co-author in co-writing systems. Then their evaluations primarily derive from two dimensions: \emph{i)} how does the human feel to collaborate with LM? (\ie ``\textbf{Perception of LM as Partner}''). This subgroup consists of both metrics of general perceptions (\eg \emph{enjoyment}, \emph{preference}, etc.), and ethical metrics (\eg \emph{stereotypical}, etc.).
Meanwhile, \emph{ii)} how credible is the LM's feedback? (\ie ``\textbf{Measure LM Feedbacks}''), which involves common metrics for a variety tasks (\eg \emph{usability}, etc.), or task-oriented metrics, like \emph{imagination} preferred by storytelling, or \emph{structural} for scientific writing, etc.

%


\vspace{2pt}
\noindent \hspace*{.0em}\img{figures/num2.pdf}\hspace*{-.1em} \textbf{Evaluating dynamic interaction trace.} These metrics focus on evaluating the dynamic change of interactions along iterative writing process. 
We identify three dimensions for evaluations. First, when human iteratively updates the artifacts, ``\textbf{Iterative Interaction Change}'' subgroup aims to compare metrics between multiple iterations, such as measuring human understanding on LM \emph{before} start writing and when almost \emph{finish} the article).
Also, we cover ``\textbf{Span Interactive Flow}'' subgroup here to assess metrics that need to observe spanning multiple artifact versions (\eg \emph{consistency}, \emph{learning curve}).
Besides, the responding time affects user experience in the interactive systems. We thus include ``\textbf{Efficiency}'' metrics (\eg \emph{latency}, \emph{incorporation rate}) to assess the process.


\vspace{2pt}
\noindent \hspace*{.0em}\img{figures/num3.pdf}\hspace*{-.1em} \textbf{Evaluating writing artifact.} These metrics gauge the content of the final writing artifact that human and LM jointly accomplish.
We broadly divide these metrics into ``\textbf{Measure Writing Artifacts}'', where we can compare written artifacts with ground-truth articles (\eg using \emph{Jaccard similarity}) or recruit external experts for evaluation when without ground truth, and ``\textbf{Perception from Writer}'' dimensions, where writers provide subjective feedback on their outputs (\eg \emph{satisfaction} or \emph{ownership}).

\vspace{-3pt}
\section{Case Study: \system for Co-writing}
\label{sec:case}

We consider \system as a framework for researchers to fairly evaluate and compare co-writing systems.
\system can be useful for researchers to:
{1)} 
\textbf{identify key interactive evaluation aspects} among human, LM, and writing artifact interactions during the co-writing process, 
{2)} 
\textbf{select appropriate metrics} to assess and compare the co-writing systems, 
{3)} 
\textbf{comprehensively analyze and describe} the human-LM interactive performance of the co-writing systems.
Next we present a concrete use case of re-evaluating the {Beyond Text Generation} (BTG)~\cite{beyondtextgeneration} system to showcase how to use \system for evaluation.

Suppose the researchers have built the BTG system and need to evaluate its performance in an interactive setting.
\system can help them assess the system comprehensively and compare it to existing baselines.
For example, they can use the \system framework to unpack their hypothesis into fine-grained evaluation requirements. 
A mapping may look like: ``{the BTG system can help humans write better articles} (\ie \emph{aspect3: evaluate writing artifact}) {by enabling better human-LM interactions} (\ie \emph{aspect1: evaluate human-LM interaction}) {in efficient ways} (\ie \emph{aspect2: evaluate dynamic interaction trace}).
With these aspects in mind, they can then dive into each aspect to select the appropriate metrics to support this statement.
For instance, they can assess ``writers' perceptions of LM'' by choosing \emph{enjoyable}, \emph{preference} metrics, and ``how writers think about LM's feedbacks'' with \emph{usabiity}, \emph{effective}, \emph{coherence} metrics, etc.
Also, they can assess the dynamic interaction efficiency by analyzing the objective logging data (\eg \emph{incorporation rate}, \emph{latency}).
The final writing article can also be rated with a set of measures (\eg \emph{external expert review}, \emph{quality}, \emph{satisfaction}, \emph{ownership}, etc.).
Following the evaluation methods presented in Appendix~\ref{appdx:methods} (which captures the most common methods used for state-of-the-art system evaluations), they can map the metrics to actual user study designs.
%
Note that the researchers apply all measurements to both the proposed BTG system and the baselines, and ideally with the same group of users, so that they can compare the performance of co-writing systems in a fair manner.

\vspace{-3pt}
\section{{\large Conclusion}}
\label{sec:conclusion}

This work present \system: a human-centered framework to evaluate human-LM interactive co-writing systems.
It provides a thinking tool for researchers to design comprehensive interaction evaluations and analyses.
We further feature a use case study introducing how to use \system step-by-step for fair evaluations.
%
%


\section{{\large Acknowledgement}}
\label{sec:acknowledgement}
We especially \textbf{Thank Dr. Ting-Hao 'Kenneth' Huang} for his insightful feedback, valuable support, and co-organizing the In2Writing Workshops!

%
%
\bibliographystyle{ACM-Reference-Format}
\bibliography{papers}


\begin{thebibliography}{12}


\ifx \showCODEN    \undefined \def \showCODEN     #1{\unskip}     \fi
\ifx \showDOI      \undefined \def \showDOI       #1{#1}\fi
\ifx \showISBNx    \undefined \def \showISBNx     #1{\unskip}     \fi
\ifx \showISBNxiii \undefined \def \showISBNxiii  #1{\unskip}     \fi
\ifx \showISSN     \undefined \def \showISSN      #1{\unskip}     \fi
\ifx \showLCCN     \undefined \def \showLCCN      #1{\unskip}     \fi
\ifx \shownote     \undefined \def \shownote      #1{#1}          \fi
\ifx \showarticletitle \undefined \def \showarticletitle #1{#1}   \fi
\ifx \showURL      \undefined \def \showURL       {\relax}        \fi
\providecommand\bibfield[2]{#2}
\providecommand\bibinfo[2]{#2}
\providecommand\natexlab[1]{#1}
\providecommand\showeprint[2][]{arXiv:#2}

\bibitem[Akoury et~al\mbox{.}(2020)]%
        {akoury-etal-2020-storium}
\bibfield{author}{\bibinfo{person}{Nader Akoury}, \bibinfo{person}{Shufan
  Wang}, \bibinfo{person}{Josh Whiting}, \bibinfo{person}{Stephen Hood},
  \bibinfo{person}{Nanyun Peng}, {and} \bibinfo{person}{Mohit Iyyer}.}
  \bibinfo{year}{2020}\natexlab{}.
\newblock \showarticletitle{{STORIUM}: {A} {D}ataset and {E}valuation
  {P}latform for {M}achine-in-the-{L}oop {S}tory {G}eneration}. In
  \bibinfo{booktitle}{\emph{Proceedings of the 2020 Conference on Empirical
  Methods in Natural Language Processing (EMNLP)}}.
  \bibinfo{publisher}{Association for Computational Linguistics},
  \bibinfo{address}{Online}, \bibinfo{pages}{6470--6484}.
\newblock
\urldef\tempurl%
\url{https://doi.org/10.18653/v1/2020.emnlp-main.525}
\showDOI{\tempurl}


\bibitem[Chen et~al\mbox{.}(2019)]%
        {chen2019gmail}
\bibfield{author}{\bibinfo{person}{Mia~Xu Chen}, \bibinfo{person}{Benjamin~N
  Lee}, \bibinfo{person}{Gagan Bansal}, \bibinfo{person}{Yuan Cao},
  \bibinfo{person}{Shuyuan Zhang}, \bibinfo{person}{Justin Lu},
  \bibinfo{person}{Jackie Tsay}, \bibinfo{person}{Yinan Wang},
  \bibinfo{person}{Andrew~M Dai}, \bibinfo{person}{Zhifeng Chen},
  {et~al\mbox{.}}} \bibinfo{year}{2019}\natexlab{}.
\newblock \showarticletitle{Gmail smart compose: Real-time assisted writing}.
  In \bibinfo{booktitle}{\emph{Proceedings of the 25th ACM SIGKDD International
  Conference on Knowledge Discovery \& Data Mining}}.
  \bibinfo{pages}{2287--2295}.
\newblock


\bibitem[Chowdhery et~al\mbox{.}(2022)]%
        {palm}
\bibfield{author}{\bibinfo{person}{Aakanksha Chowdhery},
  \bibinfo{person}{Sharan Narang}, \bibinfo{person}{Jacob Devlin},
  \bibinfo{person}{Maarten Bosma}, \bibinfo{person}{Gaurav Mishra},
  \bibinfo{person}{Adam Roberts}, \bibinfo{person}{Paul Barham},
  \bibinfo{person}{Hyung~Won Chung}, \bibinfo{person}{Charles Sutton},
  \bibinfo{person}{Sebastian Gehrmann}, {et~al\mbox{.}}}
  \bibinfo{year}{2022}\natexlab{}.
\newblock \showarticletitle{Palm: Scaling language modeling with pathways}.
\newblock \bibinfo{journal}{\emph{arXiv preprint arXiv:2204.02311}}
  (\bibinfo{year}{2022}).
\newblock


\bibitem[Dang et~al\mbox{.}(2022)]%
        {beyondtextgeneration}
\bibfield{author}{\bibinfo{person}{Hai Dang}, \bibinfo{person}{Karim
  Benharrak}, \bibinfo{person}{Florian Lehmann}, {and} \bibinfo{person}{Daniel
  Buschek}.} \bibinfo{year}{2022}\natexlab{}.
\newblock \showarticletitle{Beyond Text Generation: Supporting Writers with
  Continuous Automatic Text Summaries}. In
  \bibinfo{booktitle}{\emph{Proceedings of the 35th Annual ACM Symposium on
  User Interface Software and Technology}}. \bibinfo{pages}{1--13}.
\newblock


\bibitem[Lee et~al\mbox{.}(2022)]%
        {lee2022evaluating}
\bibfield{author}{\bibinfo{person}{Mina Lee}, \bibinfo{person}{Megha
  Srivastava}, \bibinfo{person}{Amelia Hardy}, \bibinfo{person}{John
  Thickstun}, \bibinfo{person}{Esin Durmus}, \bibinfo{person}{Ashwin
  Paranjape}, \bibinfo{person}{Ines Gerard-Ursin}, \bibinfo{person}{Xiang~Lisa
  Li}, \bibinfo{person}{Faisal Ladhak}, \bibinfo{person}{Frieda Rong},
  {et~al\mbox{.}}} \bibinfo{year}{2022}\natexlab{}.
\newblock \showarticletitle{Evaluating Human-Language Model Interaction}.
\newblock \bibinfo{journal}{\emph{arXiv preprint arXiv:2212.09746}}
  (\bibinfo{year}{2022}).
\newblock


\bibitem[Miles and Huberman(1994)]%
        {miles1994qualitative}
\bibfield{author}{\bibinfo{person}{Matthew~B Miles} {and}
  \bibinfo{person}{A~Michael Huberman}.} \bibinfo{year}{1994}\natexlab{}.
\newblock \bibinfo{booktitle}{\emph{Qualitative data analysis: An expanded
  sourcebook}}.
\newblock \bibinfo{publisher}{sage}.
\newblock


\bibitem[Mirowski et~al\mbox{.}(2022)]%
        {dramatron}
\bibfield{author}{\bibinfo{person}{Piotr Mirowski}, \bibinfo{person}{Kory~W
  Mathewson}, \bibinfo{person}{Jaylen Pittman}, {and} \bibinfo{person}{Richard
  Evans}.} \bibinfo{year}{2022}\natexlab{}.
\newblock \showarticletitle{Co-writing screenplays and theatre scripts with
  language models: An evaluation by industry professionals}.
\newblock \bibinfo{journal}{\emph{arXiv preprint arXiv:2209.14958}}
  (\bibinfo{year}{2022}).
\newblock


\bibitem[Ouyang et~al\mbox{.}(2022)]%
        {ouyang2022training}
\bibfield{author}{\bibinfo{person}{Long Ouyang}, \bibinfo{person}{Jeff Wu},
  \bibinfo{person}{Xu Jiang}, \bibinfo{person}{Diogo Almeida},
  \bibinfo{person}{Carroll~L Wainwright}, \bibinfo{person}{Pamela Mishkin},
  \bibinfo{person}{Chong Zhang}, \bibinfo{person}{Sandhini Agarwal},
  \bibinfo{person}{Katarina Slama}, \bibinfo{person}{Alex Ray},
  {et~al\mbox{.}}} \bibinfo{year}{2022}\natexlab{}.
\newblock \showarticletitle{Training language models to follow instructions
  with human feedback}.
\newblock \bibinfo{journal}{\emph{arXiv preprint arXiv:2203.02155}}
  (\bibinfo{year}{2022}).
\newblock


\bibitem[Rae et~al\mbox{.}(2021)]%
        {rae2021scaling}
\bibfield{author}{\bibinfo{person}{Jack~W Rae}, \bibinfo{person}{Sebastian
  Borgeaud}, \bibinfo{person}{Trevor Cai}, \bibinfo{person}{Katie Millican},
  \bibinfo{person}{Jordan Hoffmann}, \bibinfo{person}{Francis Song},
  \bibinfo{person}{John Aslanides}, \bibinfo{person}{Sarah Henderson},
  \bibinfo{person}{Roman Ring}, \bibinfo{person}{Susannah Young},
  {et~al\mbox{.}}} \bibinfo{year}{2021}\natexlab{}.
\newblock \showarticletitle{Scaling language models: Methods, analysis \&
  insights from training gopher}.
\newblock \bibinfo{journal}{\emph{arXiv preprint arXiv:2112.11446}}
  (\bibinfo{year}{2021}).
\newblock


\bibitem[Rezwana and Maher(2022)]%
        {rezwana2022designing}
\bibfield{author}{\bibinfo{person}{Jeba Rezwana} {and}
  \bibinfo{person}{Mary~Lou Maher}.} \bibinfo{year}{2022}\natexlab{}.
\newblock \showarticletitle{Designing Creative AI Partners with COFI: A
  Framework for Modeling Interaction in Human-AI Co-Creative Systems}.
\newblock \bibinfo{journal}{\emph{ACM Transactions on Computer-Human
  Interaction}} (\bibinfo{year}{2022}).
\newblock


\bibitem[Singh et~al\mbox{.}(2022)]%
        {integrativeheaps}
\bibfield{author}{\bibinfo{person}{Nikhil Singh}, \bibinfo{person}{Guillermo
  Bernal}, \bibinfo{person}{Daria Savchenko}, {and} \bibinfo{person}{Elena~L
  Glassman}.} \bibinfo{year}{2022}\natexlab{}.
\newblock \showarticletitle{Where to hide a stolen elephant: Leaps in creative
  writing with multimodal machine intelligence}.
\newblock \bibinfo{journal}{\emph{ACM Transactions on Computer-Human
  Interaction}} (\bibinfo{year}{2022}).
\newblock


\bibitem[Yuan et~al\mbox{.}(2022)]%
        {wordcraft}
\bibfield{author}{\bibinfo{person}{Ann Yuan}, \bibinfo{person}{Andy Coenen},
  \bibinfo{person}{Emily Reif}, {and} \bibinfo{person}{Daphne Ippolito}.}
  \bibinfo{year}{2022}\natexlab{}.
\newblock \showarticletitle{Wordcraft: story writing with large language
  models}. In \bibinfo{booktitle}{\emph{27th International Conference on
  Intelligent User Interfaces}}. \bibinfo{pages}{841--852}.
\newblock


\end{thebibliography}

\newpage

\appendix

\section{{\large Appendix}}
\label{sec:appendix}

\subsection{\system Metric Details}
\label{appdx:metrics}

We list the practical evaluation metrics of \system with details, including Interaction Aspects, Subgouprs, Metrics, Measure Questions, and References in Table~\ref{tab:framework}.

\subsection{Evaluation Methods in User Studies for Co-writing Systems}
\label{appdx:methods}

\noindent 
We summarize a list of evaluation methods that are commonly used in studying co-writing systems~\cite{integrativeheaps,dramatron,wordcraft} for the future work reference.
This summary aims to serve the purpose of providing inspirations and benchmarks for future co-writing system work to design and compare user study evaluations. 

(a) \textbf{Coding users' think aloud transcripts.} Researchers can encourage the participants to articulate their thinking during interacting with the systems, such as \emph{why they decided to use this prompt for querying LM but not others}, etc.
After finishing the study, researchers can convert the video/radio into transcripts and code the transcripts for further data analysis. 
Some common qualitative data analysis approaches include thematic analysis, content coding, and topic modeling~\cite{miles1994qualitative}, etc.

(b) \textbf{Coding researchers' observation notes.}
During the user studies, the researchers can also record their observations for later analysis. For instance, they can pre-design a set of topics (\eg user's emotion change, etc.) that are important to answer their research questions, and pay close attention to these topics during the studies.

(c) \textbf{Coding (semi-)structured interview transcripts.}
Prior studies also commonly leverage (semi-)structured interviews to elicit users' experience and feedback on using the systems. Researchers can design effective interview questions and invoke participants' answers, to better support the research arguments.

(d) \textbf{Questionnaires or Surveys.}
Previous studies also frequently design surveys, which include questions such as N-five-point Likert ratings, single choice or multiple choice questions, etc. These surveys can provide more accurate measures on user's assessment.

(e) \textbf{Interaction data logs.} The data logs during interaction can provide more objective analysis on user behaviors. Typical interaction data logs for co-writing systems involve artifact submission count, prompt request frequency, latency, etc.

(f) \textbf{Assessment on the written artifacts.}
These metrics aim to directly evaluate the quality of written artifacts, commonly using automatic metrics by comparing with ground truth (\eg \emph{similarity}), computing the artifact properties (\eg \emph{word count}, \emph{document length}, etc.), or having external experts to assess the outputs.


%
%

\begin{table*}[]
\footnotesize
\begin{tabular}[t]{@{} p{0.14\textwidth} | p{0.19\textwidth} | p{0.18\textwidth} | p{0.33\textwidth} | p{0.08\textwidth}  @{}}

\toprule
\textbf{Interaction Aspects}                       & \textbf{Subgroups}                               & \textbf{Metrics} & \textbf{Measure Questions} & \textbf{References} \\ \midrule
\multirow{6}{*}{
%
\makecell{\ding{202} \textbf{Evaluating Human-} \\ \textbf{-LM Interaction}}}
& \multirow{4}{*}{\makecell{\textbf{Perception of LM as Partner} \\ \textbf{[General perceptions}]}}   & 
{\cellcolor[HTML]{DAE8FC}{enjoyment}}
& \emph{I enjoyed writing the story.}  & \cite{wordcraft,dramatron}  \\ \cline{3-5}

   &   & {\cellcolor[HTML]{DAE8FC}{effort}}  & \emph{I put lots of effort into getting AI suggestions.}  &  \cite{integrativeheaps} \\ \cline{3-5}
   &   & {\cellcolor[HTML]{DAE8FC}{preference}}  & \emph{I prefer the suggestions from the AI agent.}  & \cite{dramatron}  \\ \cline{3-5}
   &   & {\cellcolor[HTML]{DAE8FC}{communicative}}  &  \emph{I can communicate well with the AI agent.}  & \cite{integrativeheaps}  \\ \cline{3-5}
   &   & {\cellcolor[HTML]{DAE8FC}{cognitive load}}   & \emph{Interacting with the AI agent requires much cognitive load. }  &  \cite{integrativeheaps} \\ \cline{3-5}
   &   &  {\cellcolor[HTML]{DAE8FC}{collaborative}} &  \emph{I feel the AI agent collaborative to work together.} &  \cite{wordcraft,integrativeheaps,dramatron} \\ \cline{3-5} 
   &   &  {\cellcolor[HTML]{DAE8FC}{ease}}  & \emph{The AI agent is easy to learn and work with.}  &  \cite{wordcraft,dramatron} \\ \cmidrule{2-5} 
   & \multirow{2}{*}{\makecell{\textbf{Perception of LM as Partner} \\ \textbf{[Ethical Metrics}]}}  & {\cellcolor[HTML]{DAE8FC}{understanding}}   & \emph{I can understand the AI agent.}  &  \cite{integrativeheaps} \\ \cline{3-5}
    &   & {\cellcolor[HTML]{DAE8FC}{literal}}   & \emph{The AI agent's suggestions are literal}  & \cite{dramatron} \\ \cline{3-5} 
    &   & {\cellcolor[HTML]{DAE8FC}{stereotypical}}   & The AI agent's suggestions are stereotypical & \cite{dramatron} \\ \cline{3-5} 
    &   &  {\cellcolor[HTML]{DAE8FC}{non-intentional}}  & The AI agent does not have intentions to generate suggstions.  & \cite{integrativeheaps}  \\ \cmidrule{2-5}
   & \multirow{2}{*}{\makecell{\textbf{Measure LM Feedbacks} \\ \textbf{[Common Metrics}]}}
   & {\cellcolor[HTML]{DAE8FC}{coherent}}
   &  \emph{The AI generations are coherent with prompts.} & \cite{integrativeheaps,dramatron}  \\ \cline{3-5}
   &   & 
   {\cellcolor[HTML]{DAE8FC}{variety}}
    & \emph{The AI agent's suggestion are various.}  & \cite{integrativeheaps,dramatron}  \\ \cline{3-5}
   &   & {\cellcolor[HTML]{DAE8FC}{helpful}}  &  \emph{I found the AI agent helpful.} &  \cite{wordcraft,integrativeheaps,dramatron} \\ \cline{3-5}
   &   & {\cellcolor[HTML]{DAE8FC}{effective}}  & \emph{The AI agent is effective at suggesting ideas.}  & \cite{wordcraft}  \\ \cline{3-5}
   &   & {\cellcolor[HTML]{DAE8FC}{repetition}}  & \emph{The AI agent generates repetitive suggestions.} & \cite{integrativeheaps}  \\ \cline{3-5}
   &   & {\cellcolor[HTML]{DAE8FC}{usability}}  & \emph{The AI agent is useful for my writing.}  & \cite{integrativeheaps}  \\ \cline{3-5}

   &   &  {\cellcolor[HTML]{DAE8FC}{combinatorial}}  & \emph{I feel the AI agent combines a broad set of information.}  & \cite{integrativeheaps}   \\ 
   \cmidrule{2-5}
   & \multirow{2}{*}{\makecell{\textbf{Measure LM Feedbacks} \\ \textbf{[Task-oriented Metrics}]}}
   &  {\cellcolor[HTML]{DAE8FC}{uniqueness}}  
   & \emph{The AI's suggestions are unique.}  & \cite{wordcraft,integrativeheaps,dramatron}  \\ \cline{3-5} 
   &   & {\cellcolor[HTML]{DAE8FC}{reality}}   & emph{The AI's suggestion aligns with common sense.}  & \cite{integrativeheaps,dramatron}  \\ \cline{3-5} 
   &   & {\cellcolor[HTML]{DAE8FC}{novelty}}   &  \emph{The AI agent often generates unexpected suggestions.} & \cite{integrativeheaps}  \\ \cline{3-5}
   &  & {\cellcolor[HTML]{DAE8FC}{freedom}}   &  \emph{I feel the AI agent can express freely.} & \cite{integrativeheaps}  \\ \cline{3-5} 
   &   & {\cellcolor[HTML]{DAE8FC}{structural}}   & \emph{The AI suggestions are structural.}  &  \cite{dramatron} \\ \cline{3-5} 
   &   & {\cellcolor[HTML]{DAE8FC}{imagination}}   &  \emph{I feel the AI agent has much imagination.} & \cite{integrativeheaps}  \\ \cline{3-5}
   &   & {\cellcolor[HTML]{DAE8FC}{unexpected}}   & \emph{The AI suggestions are often unexpected to me.}  & \cite{integrativeheaps,dramatron}  \\ 
   \midrule
\multirow{6}{*}{\makecell{\ding{203} \textbf{Evaluating Dynamic} \\ \textbf{Interaction Trace}}}    & 
\multirow{2}{*}{\makecell{\textbf{Iterative Interaction Change} \\ \textbf{[diff from prior knowledge]}}} 
& 
{\cellcolor[HTML]{DAE8FC}{different from initial expectation}}  & \emph{what's the difference before the initial expression.}  &  \cite{integrativeheaps} \\ \cline{3-5} 
   &   & {\cellcolor[HTML]{DAE8FC}{adjust to prior expectation }} &  \emph{I adjusted my expectation to prior ones.} &  \cite{integrativeheaps} \\ \cmidrule{2-5}
   &
   \multirow{2}{*}{\makecell{\textbf{Iterative Interaction Change} \\ \textbf{[diff from prev. interact. states]}}} 
   %
   %
   %
   & 
   {\cellcolor[HTML]{DAE8FC}{dynamics of suggestion integration}} &  \emph{how does the suggestion integration change dynamically.} &  \cite{integrativeheaps} \\ \cmidrule{2-5} 
   %
   %
   %
   %
   %
   & \multirow{2}{*}{\textbf{Span Interactive Flow}}     & {\cellcolor[HTML]{DAE8FC}{learning curve}}  & \emph{I can learn to use this system quickly.}  &  \cite{integrativeheaps} \\ \cline{3-5} 
   &   &  {\cellcolor[HTML]{DAE8FC}{consistency}} &  \emph{The AI generate consistent suggestions along interaction.} & \cite{integrativeheaps,dramatron}  \\ \cline{3-5}
   &   &  {\cellcolor[HTML]{DAE8FC}{flow and ordering}} & \emph{the flow and ordering of co-writing are smooth.}  &  \cite{integrativeheaps,dramatron} \\ \cmidrule{2-5}
   & \multirow{2}{*}{\textbf{Efficiency}}  & {\cellcolor[HTML]{FDE0A0}{latency}}  &  \emph{The elapsed time from human request to AI response.} & \cite{beyondtextgeneration}  \\ \cline{3-5} 
   &   & {\cellcolor[HTML]{FDE0A0}{incorporation rate}}  &  \emph{The rate of incorporating AI suggestions.} & \cite{beyondtextgeneration}   \\ \cline{3-5} 
   &   & {\cellcolor[HTML]{FDE0A0}{request count}}  &  \emph{The count of human requests.} &  \cite{beyondtextgeneration}  \\ \cline{3-5} 
   &   & {\cellcolor[HTML]{FDE0A0}{time considering suggestions}}  & \emph{The average time for human to consider AI suggestions.}  &  \cite{beyondtextgeneration}  \\ \cline{3-5} 
   &   & {\cellcolor[HTML]{FDE0A0}{\#accepted suggestions}}  &  \emph{The count of accepted AI suggestions.} & \cite{beyondtextgeneration}   \\ \cline{3-5} 
   &   & {\cellcolor[HTML]{FDE0A0}{time to complete}}  & \emph{The elapsed time for human to complete the task.}  &  \cite{beyondtextgeneration}  \\ \midrule
\multirow{6}{*}{\makecell{\ding{204} \textbf{Evaluating Writing} \\ \textbf{Artifact}}} & 
%
%
%
\multirow{2}{*}{\makecell{\textbf{Measure Writing Artifacts } \\ \textbf{[With Ground Truth]}}}
&
 {\cellcolor[HTML]{FDE0A0}{word edit distance}}  &  \emph{The word edit distance between prior- and post- articles.} &  \cite{dramatron} \\ \cline{3-5} 
   &   & {\cellcolor[HTML]{FDE0A0}{lemma-based Jaccard similarity}} & \emph{The similarity of ground truth and outcome article.}  &   \cite{dramatron} \\ \cline{3-5}
   &   & {\cellcolor[HTML]{FDE0A0}{document length difference}}  & \emph{The difference between prior- and post- articles.}  & \cite{dramatron}  \\ \cmidrule{2-5} 

   &
\multirow{2}{*}{\makecell{\textbf{Measure Writing Artifacts } \\ \textbf{[No Ground Truth]}}}
&  {\cellcolor[HTML]{DAE8FC}{outcome creativity}}  & \emph{The article I wrote with AI is creative.}  & \cite{integrativeheaps,dramatron}  \\ \cline{3-5} 
   &   & {\cellcolor[HTML]{DAE8FC}{quality}}  & \emph{The outcome article is high-quality.}  &  \cite{lee2022evaluating} \\ \cline{3-5} 
   &   & {\cellcolor[HTML]{DAE8FC}{external expert evaluation}}  & \emph{The external experts assess the writing artifacts.}  &  \cite{lee2022evaluating} \\ \cline{3-5} 
   &   & {\cellcolor[HTML]{FDE0A0}{word count}}  & \emph{The total words count of the outcome article.}  & \cite{beyondtextgeneration}  \\ \cline{3-5}
   &   & {\cellcolor[HTML]{FDE0A0}{open-source repetition scorer}}  & \emph{Computing repetition score using exisint tools.}  & \cite{dramatron}  \\ \cmidrule{2-5} 
   & 
  \multirow{2}{*}{\makecell{\textbf{Perception from Writer} \\ \textbf{[Satisfaction]}}}
   &  {\cellcolor[HTML]{DAE8FC}{writing goal}} & \emph{The outcome article reaches my writing goal.} &  \cite{beyondtextgeneration,dramatron} \\ \cline{3-5}
   &   & {\cellcolor[HTML]{DAE8FC}{pride}}  & \emph{I’m proud of the final article.}  & \cite{beyondtextgeneration,dramatron}  \\ \cline{3-5}
   &   & {\cellcolor[HTML]{DAE8FC}{satisfied}}  &  \emph{I feel satisfied with the final article.}  & \cite{dramatron}  \\ \cmidrule{2-5} 
   & 
  %
  \multirow{2}{*}{\makecell{\textbf{Perception from Writer} \\ \textbf{[Ownership]}}}
   &  {\cellcolor[HTML]{DAE8FC}{ownership}} & \emph{I feel ownership over the final article.}  &  \cite{wordcraft,integrativeheaps,dramatron} \\ \cline{3-5} 
   &  & {\cellcolor[HTML]{DAE8FC}{authorial discretion}}  &  \emph{I can decide what/how to put the AI suggestions into the article.} & \cite{integrativeheaps}  \\
   \bottomrule
\end{tabular}   
\vspace{3pt}
\caption{\label{tab:framework} The practical subgroup metrics for each evaluation aspect in \system. We demonstrate the metric details of Interaction Aspects, Subgroups, Metrics, Measure Questions, and the corresponding References.
( \colorbox[HTML]{FECC66}{\color[HTML]{4B4D4C}{\texttt{yellow}}} and \colorbox[HTML]{4C91C4}{\color[HTML]{FFFFFF}{\texttt{blue}}} indicates objective and subjective metrics, respectively.)
}
\end{table*}

\end{document}